\documentclass[aps,prl,reprint,groupedaddress]{revtex4-1}

\usepackage{graphicx}
\usepackage{dcolumn}
\usepackage{amsmath}
\usepackage{amssymb}
\usepackage{wasysym}
\usepackage{color}

\usepackage{mathtools}

\begin{document}


\title{Thermodynamic evidence for electron correlation-driven flattening of the quasiparticle bands in the high-$T_{\rm c}$ cuprates 
}


\author{N.~Harrison and M.~K.~Chan}
\affiliation{Los Alamos National Laboratory, Los Alamos, NM 87545, USA}


\date{\today}

\begin{abstract}
A flattened electronic band is one of several possible routes for increasing the strength of the pairing interactions in a superconductor. With this in mind, we show here that thermodynamic measurements of the high-$T_{\rm c}$ cuprates reveal an appreciably stronger electron correlation-driven flattening of the quasiparticle bands than has previously been indicated. Specifically, we find that thermodynamic measurements indicate an electronic entropy in excess of that that can be accounted for by the value of the universal Fermi velocity inferred from photoemission experiments. The observed band flattening implies that the Van Hove singularity features prominently in calorimetry measurements, causing it undermine prior arguments for a divergence in the renormalization of the effective mass near a critical doping $p^\ast$ based on calorimetry measurements. The band flattening is also sufficient to drive the cuprates into a strong pairing regime where the maximum transition temperature becomes constrained by phase fluctuations. %
\end{abstract}


\maketitle
%
%

The degree of flatness of the electronic bands in a material has been shown to be a chief determining factor in the realization of correlated insulating states and superconducting states with strong pairing interactions~\cite{cao2018,park2021}. Correlated insulating states are well documented in the undoped parent phases of the high temperature superconducting cuprates~\cite{keimer2015,lee2006}. There are also a number of physical phenomena in hole-doped cuprates that are consistent with strong pairing interactions~\cite{baskaran1987,emery1995,timusk1999,hufner2008,harrison2022,uemura1989,uemura1991},  suggesting therefore a possible role of phase fluctuations in constraining the maximum value of the critical temperature $T_{\rm c}$~\cite{hazra2019,shi2022,quasi2D,ries2015,li2007,guo2020}. In a tight-binding lattice, phase fluctuations are predicted to cause the upper limit for the transition temperature to be given by $k_{\rm B}T_{\rm c}/t\sim$~0.2~\cite{denteneer1993,chen1999,keller2001,toschi2005,paiva2010,pasrija2016,paiva2004}, where $t$ is the in-plane hopping. Given the large value of $t\approx$~360~meV predicted by electronic structure theory in the cuprates~\cite{matteiss1987,massidda1988,singh1992,andersen1995}, a significant band renormalization is required for the upper limit of $T_{\rm c}$ to have been reached~\cite{keimer2015}. 

At present, our understanding of the effective mass renormalization in the cuprates relies heavily on angle-resolved photoemission spectroscopy measurements~~\cite{johnson2001,byczuki2007,lanzara2001}, to the extent that it is photoemission experimental results against which all other experimental results are routinely compared~\cite{padilla2005,sebastian2012,ramshaw2015,legros2022,michon2019,zhong2022,storey2008,doiron2007,sous2022}. Photoemission experiments have reported the nodal Fermi velocity to exhibit a universal value of $v_{\rm F}\approx$~2.7~$\times$~10$^5$~ms$^{-1}$~\cite{yoshida2003,damascelli2003}, which is roughly half the velocity $v_{\rm b}\approx$~5.1~$\times$~10$^5$~ms$^{-1}$ inferred from non-interacting electronic band theory~\cite{matteiss1987,massidda1988,singh1992,andersen1995}, and corresponds to a reduced hopping of $t\approx$~190~meV~\cite{horio2018,michon2019,hopping,drozdov2018,zhong2022}. Upon taking a simple ratio of velocities, ${v_{\rm b}}/{v_{\rm F}}$, one can infer a modest enhancement of the nodal photoemission effective mass $m^\ast$ relative to the band mass $m_{\rm b}$ of ${m^\ast}/{m_{\rm b}}\approx$~1.9. 

\begin{figure}
\begin{center}
\includegraphics[width=0.95\linewidth]{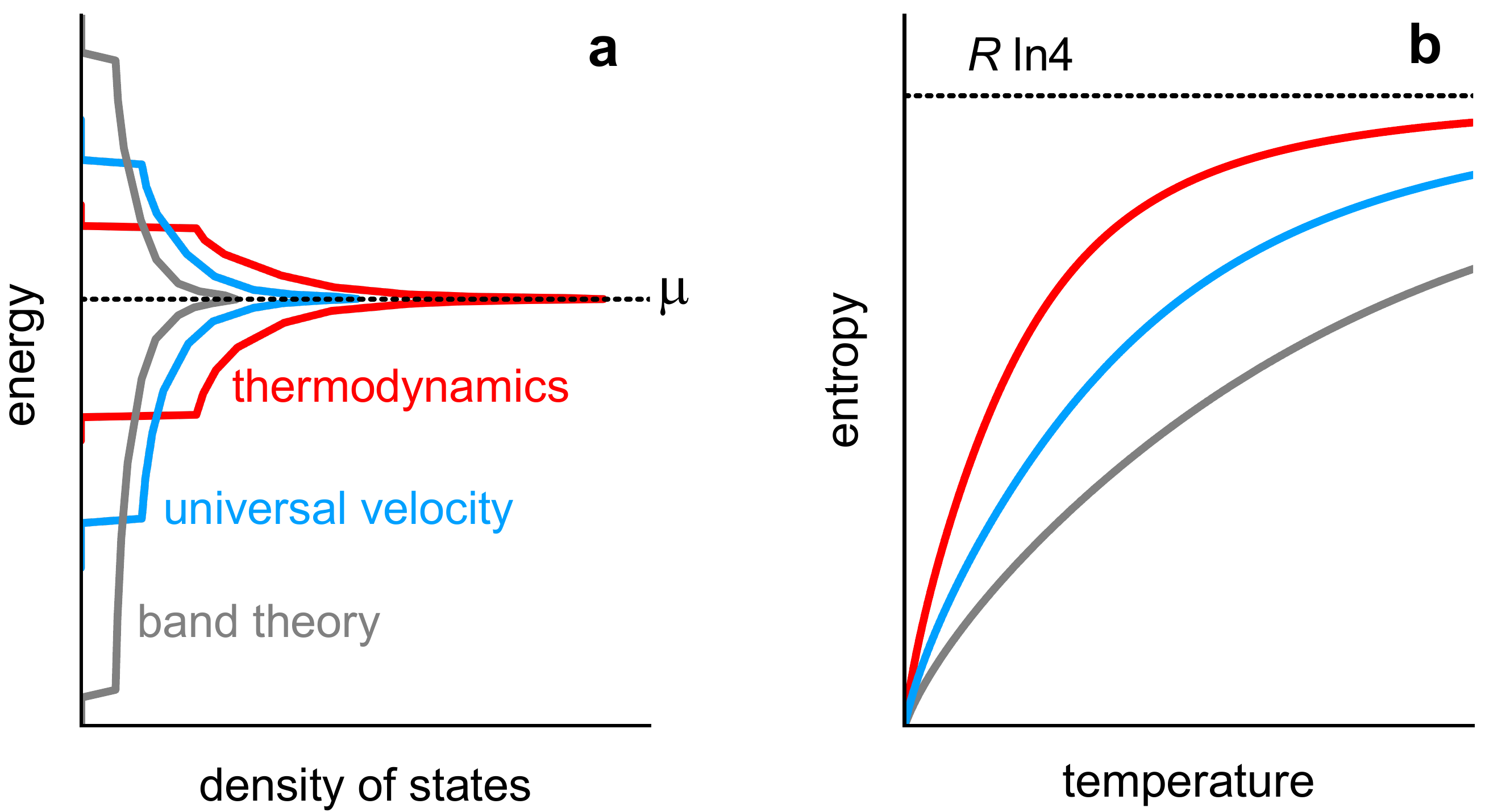}
\textsf{\caption{{\bf a}, Schematic electronic density density of states for values of the nodal velocity (or $t$) according to band theory (grey), the universal velocity inferred from angle-resolved photoemission spectroscopy measurements (blue) and that we infer from thermodynamic measurements (red). {\bf b}, Schematic of the corresponding entropy versus temperature curves.
}
\label{bandentropy}}
\end{center}
\vspace{-0.7cm}
\end{figure}

In this paper, we show that there is a significant excess electronic entropy in thermodynamic measurements that is not accounted for by the universal velocity inferred from photoemission measurements~\cite{yoshida2003}. Moreover, rather than being restricted to a singular point~\cite{mathur1998,laughlin2001,sachdev2010,shibauchi2013,coleman2005}, as has been suggested~\cite{michon2019}, the excess entropy occurs over a wide range of hole dopings.  It is most clearly evidenced by the magnitude of the electronic entropy determined from calorimetry measurements extending up to room temperature~\cite{loram2001,loram1994,loram1998}. It is also evidenced by the small values of the orbitally-averaged Fermi velocity $v_{\rm F}^{\rm orb}$ obtained from quantum oscillation measurements made at low temperatures~\cite{rourke2010,ramshaw2015,tan2015,barisic2013,chan2016,kunisada2020,oliviero2022}. Finally, we show that it can account for recent low temperature calorimetry measurements~\cite{horio2018,michon2019} without invoking quantum critical fluctuations.  We find that the excess electronic entropy in all three thermodynamic measurements can be consistently understood by considering a significantly smaller value of the hopping of $t\approx$~107~meV (see illustration in Fig.~\ref{bandentropy}), corresponding to a reduced nodal Fermi velocity of $v_{\rm F}\approx$~1.5~$\times$~10$^5$~ms$^{-1}$. We further show that this reduced hopping increases the prominence of the Van Hove singularity in calorimetry measurements, thereby weakening arguments for a divergence in the effective mass enhancement at the critical doping $p^\ast$~\cite{michon2019}. We show that the reduced hopping is also sufficient to drive the pairing into a regime where phase fluctuations become relevant in constraining $T_{\rm c}$. 

\begin{figure}
\begin{center}
\includegraphics[width=0.95\linewidth]{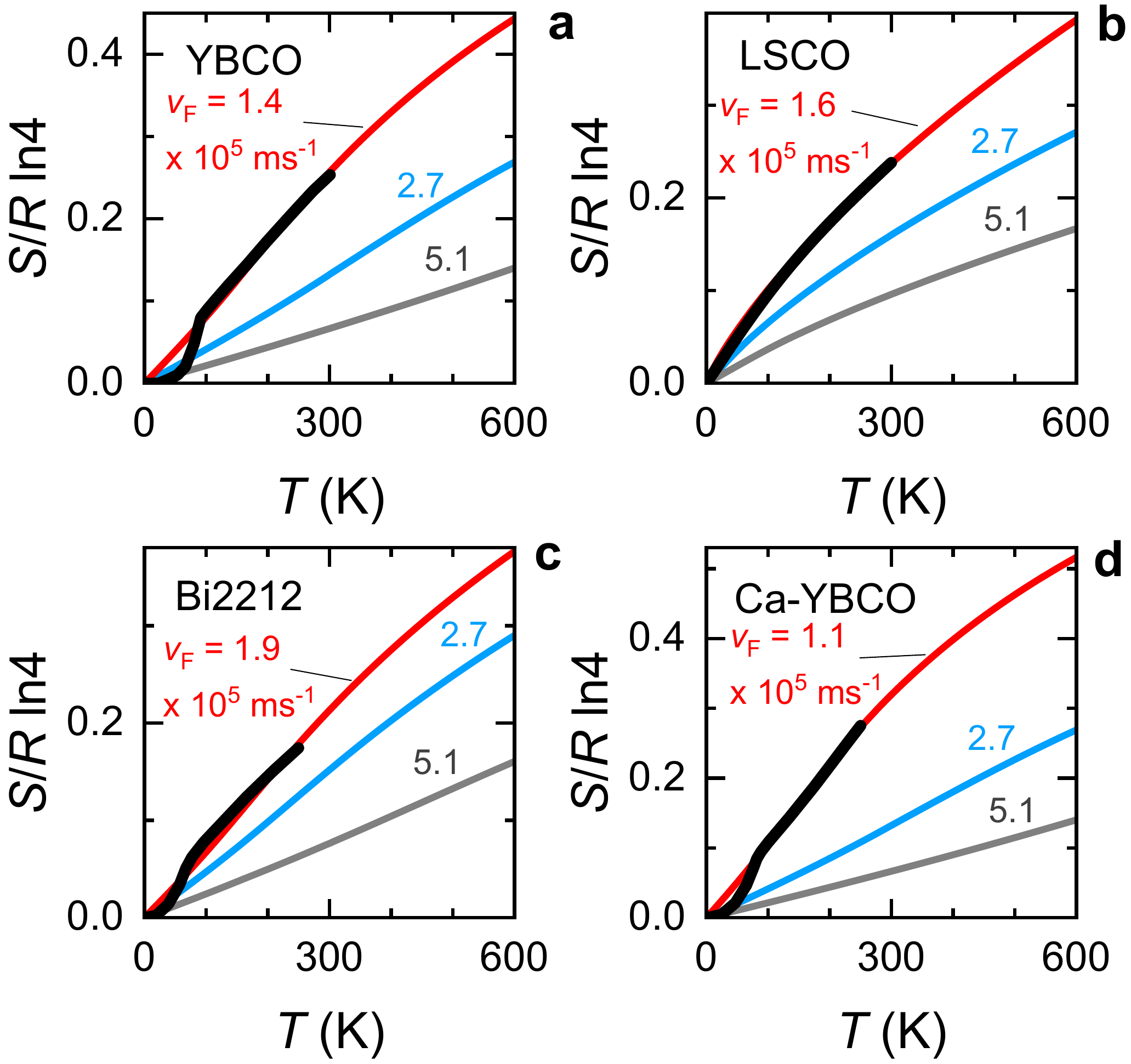}
\textsf{\caption{{\bf a}, {\bf b}, {\bf c} and {\bf d}, $S$ per planar Cu versus $T$ for YBa$_2$Cu$_3$O$_{6+x}$ (YBCO) at $p=$~0.195, La$_{2-x}$Sr$_x$CuO$_4$ (LSCO) at $p=$~0.23 Bi$_2$Sr$_2$CaCu$_2$O$_{8-\delta}$ (Bi2212) at $p=$~0.19 and 20~\%~Ca-doped YBa$_2$Cu$_3$O$_{6+x}$ (Ca-YBCO) at $p=$~0.2. After Fig.~\ref{bandentropy}, blue and grey curves correspond to $S$ versus $T$ calculated using the universal velocity (from photoemission) and the band velocity, respectively (to avoid clutter, only the first two digits of $v_{\rm F}$ are indicated). For the red curves, $t$ is adjusted (yielding $v_{\rm F}$ values indicated) to match the experimental data; their average yields $v_{\rm F}=$~(1.5~$\pm$~0.2)~$\times$~10$^5$~ms$^{-1}$.
}
\label{entropy}}
\end{center}
\vspace{-0.7cm}
\end{figure}

\begin{figure}
\begin{center}
\includegraphics[width=0.95\linewidth]{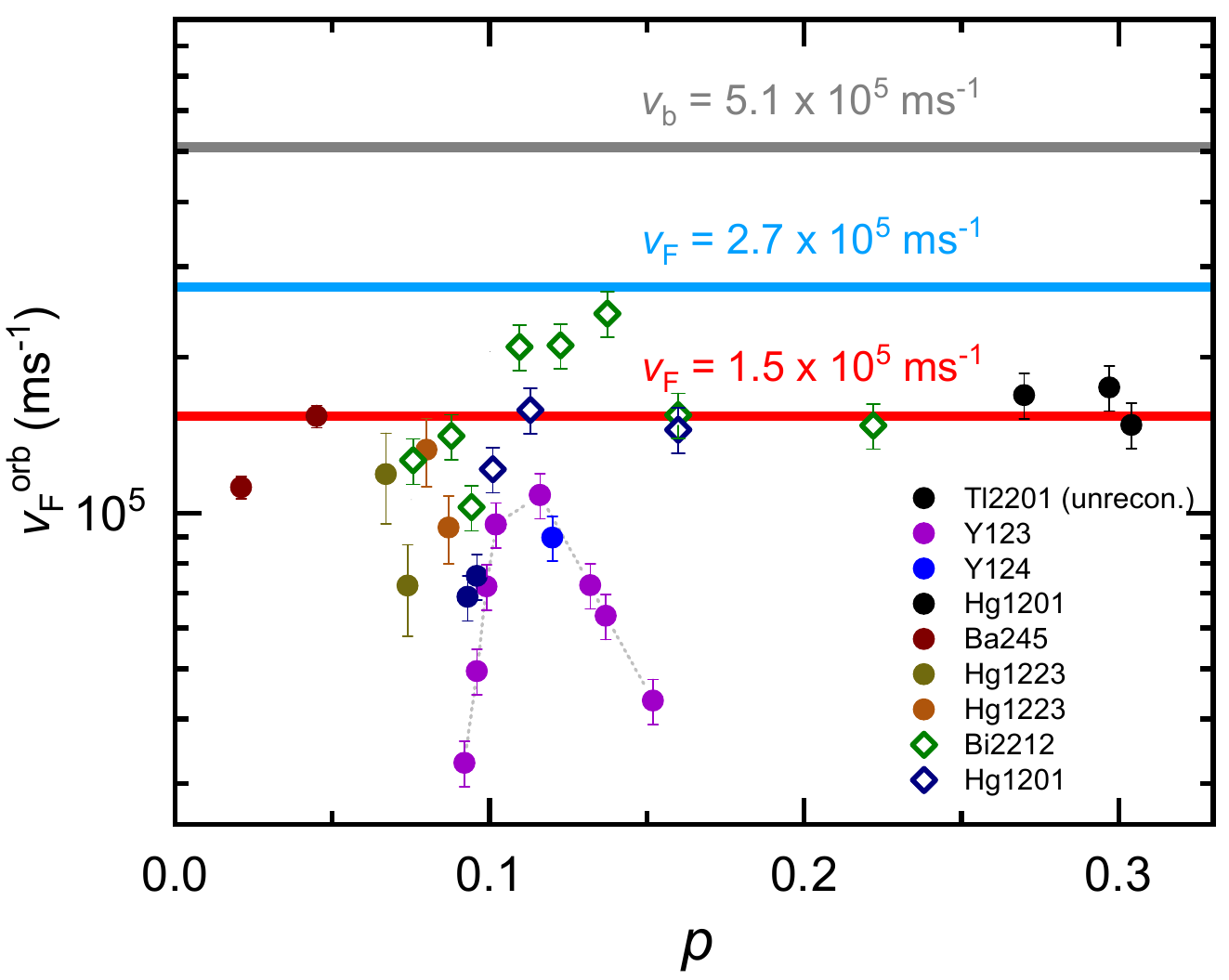}
\textsf{\caption{Orbitally-averaged $v_{\rm F}$ ($\circ$ symbols) for cuprates in which quantum oscillations have been observed, including overdoped Tl$_2$Ba$_2$CuO$_6$ (Tl2201)~\cite{rourke2010}, and underdoped YBa$_2$Cu$_3$O$_{6+x}$ (Y123)~\cite{ramshaw2015}, YBa$_2$Cu$_4$O$_8$ (Y124)~\cite{tan2015}, HgBa$_2$CuO$_{4+\delta}$ (Hg1201)~\cite{barisic2013,chan2016}, Ba$_2$Ca$_4$Cu$_5$O$_{10}$(F,O)$_2$ (Ba245)~\cite{kunisada2020} and HgBa$_2$Ca$_2$Cu$_3$O$_{8+\delta}$ (Hg1223)~\cite{oliviero2022}. These are compared against band theory (grey line), the photoemission universal velocity~\cite{yoshida2003} (blue line), and the thermodynamic $v_{\rm F}$ from Fig.~\ref{entropy} (red line). Also shown are $v_{\rm F}$ values determined from high resolution laser photoemission measurements~\cite{vishik2010,plumb2010,wreedhar2020} ($\diamond$ symbols).
}
\label{orbitalvelocity}}
\end{center}
\vspace{-0.7cm}
\end{figure}

\begin{figure}
\begin{center}
\includegraphics[width=0.95\linewidth]{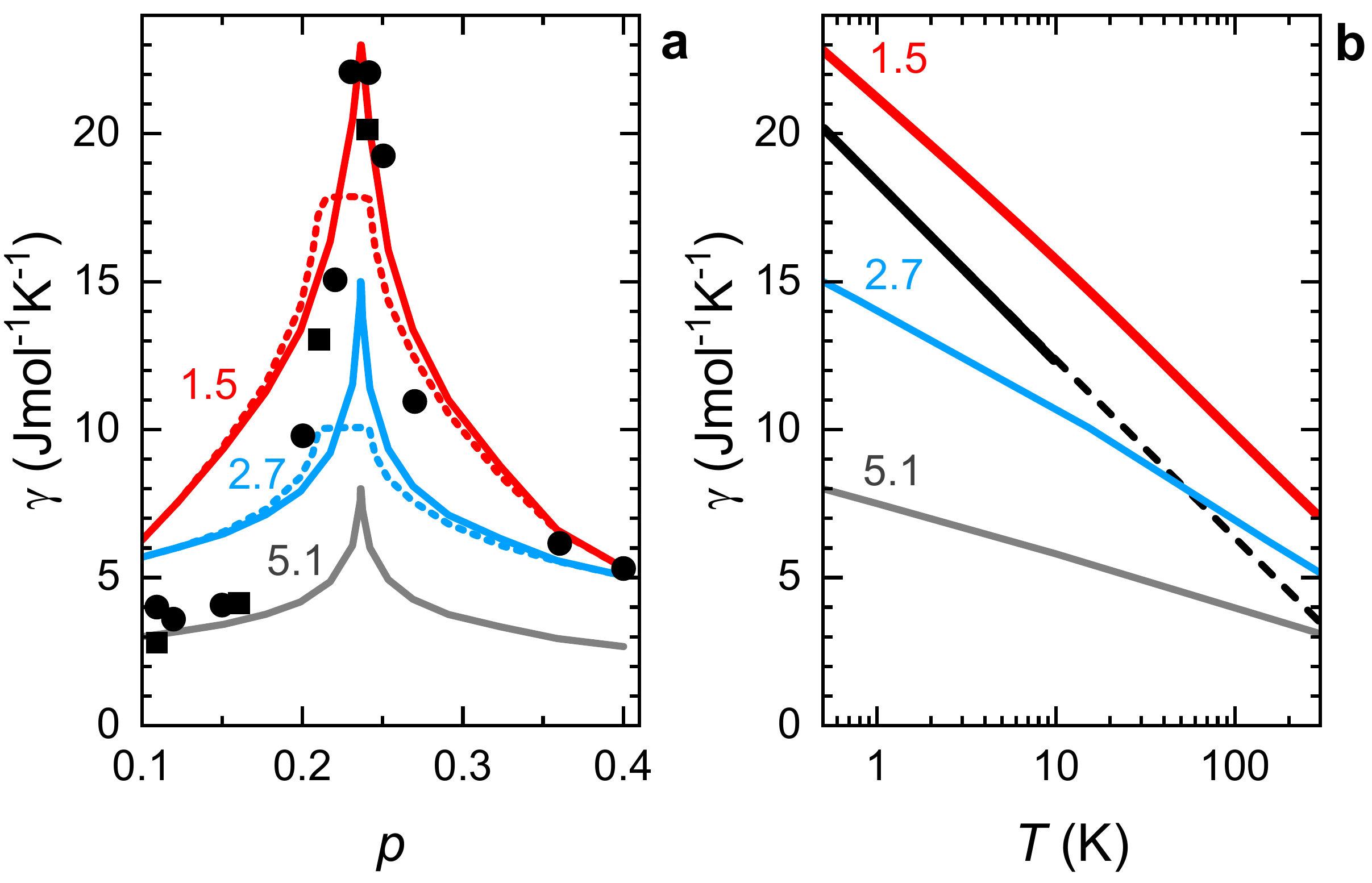}
\textsf{\caption{{\bf a}, Measured values of $\gamma$ versus $p$ at $T=$~0.5~K for  La$_{2-y-x}$Nd$_y$Sr$_x$CuO$_4$ (Nd-LSCO)~\cite{michon2019} (black $\circ$), La$_{2-y-x}$Eu$_y$Sr$_x$CuO$_4$ (Eu-LSCO) (black $\square$). 
These are compared against predictions from band theory (grey curve), the photoemission universal velocity (blue curve) and that corresponding to $v_{\rm F}$ from thermodynamic measurements averaged from Fig.~\ref{entropy} (red curve). To avoid clutter, only the first two digits of $v_{\rm F}$ are indicated. Dotted curves include the effect of $c$-axis hopping ($t_z=0.07t$~\cite{horio2018}). For the red curve, $v_{\rm F}$ (via $t$) is adjusted to account for the smooth variation of $S$ with $p$ observed at 300~K in LSCO~\cite{loram2001}. {\bf b}, Representation of the measured (solid black line) and extrapolated (dotted black line) $\gamma$ versus $\ln T$ curve in Eu-LSCO~\cite{michon2019}. Red, blue and grey lines show $\gamma$ versus $T$ at the Van Hove singularity ($p\approx$~0.23) using the same velocity values as in ({\bf a}).
}
\label{lowtcalorimetry}}
\end{center}
\vspace{-0.7cm}
\end{figure}

Figures~\ref{entropy}, \ref{orbitalvelocity} and \ref{lowtcalorimetry} show the thermodynamic evidence for flatter quasiparticle bands arising from stronger electronic correlations in the cuprates. To assist in the interpretation of the experimental data, we first clarify the relationships between $t$ and the values of Sommerfeld coefficient $\gamma$, the Fermi velocity $v_{\rm F}$ and the entropy $S$. They can be understood by considering the standard tight binding approximation $\epsilon_{\bf k}=-2t(\cos ak_x+\cos ak_y)+4t^\prime\cos ak_x\cos ak_y-2t^{\prime\prime}(\cos2ak_x+\cos2ak_y)-\mu$ to the electronic dispersion~\cite{drozdov2018,tightbindingparameters}, where $t^\prime$ and $t^{\prime\prime}$ are higher order hopping parameters and $a$ is the in-plane lattice parameter. 
The magnitude of the nodal Fermi velocity is given by the gradient~\cite{ashcroft1976} 
\begin{equation}\label{fermivelocity}
v_{\rm F}=\Big|\frac{1}{\hbar}\nabla_{\bf k}\varepsilon_{\bf k}\Big|
\end{equation}
and by substituting $k_x=k_y=k_{\rm F}/\sqrt{2}$ for the nodal direction, where $k_{\rm F}$ here refers to the nodal Fermi wave vector. 
The Sommerfeld coefficient at a given temperature is obtained from the integral~\cite{ashcroft1976,harrison2022i} 
\begin{equation}\label{sommerfeld}
\gamma T={N_{\rm A}}\int_{-\infty}^\infty\varepsilon D(\varepsilon){\partial{\rm f}^\prime(\varepsilon-\mu)}{\rm d}\varepsilon
\end{equation}
over the product of the electronic density of states $D(\varepsilon)=\frac{2na^2}{\pi^2}\frac{\partial}{\partial\varepsilon_{\bf k}}\int^{\pi/a}_0|{\rm Re}(k_y)|{\rm d}k_x$ and the derivative in temperature ${\rm f}^\prime(\varepsilon-\mu)$ of the Fermi-Dirac distribution function. Here, $N_{\rm A}$ is Avogadro's number, $n$ is the number of CuO$_2$ planes, while $\mu$ is the chemical potential, which is adjusted to produce a band filling consistent with experiment.
Finally, the entropy is given by the integral 
\begin{equation}\label{entropyintegration}
S=\int_0^\infty\gamma{\rm d}T.
\end{equation}
As illustrated in Fig.~\ref{bandentropy}, a smaller $t$ produces a flatter quasiparticle band with a larger electronic density of states, for which the entropy rises towards the saturation value of $R\ln4$ more rapidly with increasing temperature (where $R=N_{\rm A}k_{\rm B}$ is the gas constant), yielding the approximate proportionalities $\gamma\propto{1}/{v_{\rm F}}\propto S\propto t$. 

We turn first to the entropy as a function of temperature determined from calorimetry measurements extending from low temperature up to room temperature in Fig.~\ref{entropy}~\cite{loram2001,loram1994,loram1998}. The experimental evidence for flatter quasiparticle bands is directly apparent from the fact that the measured entropy $S$ curves (plotted in black) for slightly overdoped YBCO, LSCO, Bi2212 and Ca-YBCO lie significantly above those calculated (blue curves) using Equations~(\ref{fermivelocity}), (\ref{sommerfeld}) and (\ref{entropyintegration}) and using the universal $v_{\rm F}=$~2.7~$\times$~10$^5$~ms$^{-1}$ from photoemission experiments~\cite{yoshida2003}. 

Turning next to the results of magnetic quantum oscillation measurements in Fig.~\ref{orbitalvelocity}, more strongly renormalized quasiparticle bands are suggested by our finding that the orbitally-averaged values of the Fermi velocity $v^{\rm orb}_{\rm F}$ (circle symbols) fall significantly below the universal $v_{\rm F}$ (horizontal blue line) inferred from photoemission measurements~\cite{yoshida2003}. The orbitally-averaged velocity is estimated using $v^{\rm orb}_{\rm F}=\sqrt{2e\hbar F}/m^\ast$, where $k_{\rm F}^{\rm orb}=\sqrt{2eF/\hbar}$ is the orbitally-averaged Fermi radius, $F$ is the quantum oscillation frequency~\cite{shoenberg1984} and $m^\ast$ is the quasiparticle effective mass obtained from the temperature-dependence of the oscillations~\cite{rourke2010,ramshaw2015,tan2015,barisic2013,chan2016,kunisada2020,oliviero2022}. While some degree of reduction of $v_{\rm F}^{\rm orb}$ relative to $v_{\rm F}$ can be attributed to Fermi surface reconstruction and other factors~\cite{otherfactors}, this is not expected to be the case for the large unreconstructed Fermi surface of the overdoped cuprate Tl2212~\cite{rourke2010}. 

Finally, turning to recent low temperature calorimetry measurements in Fig.~\ref{lowtcalorimetry}, we find stronger electronic correlations to be revealed by the fact that the experimental values of $\gamma$ (black symbols) in Nd-LSCO and Eu-LSCO~\cite{michon2019} exceed those (solid blue curve) calculated using Equations~(\ref{fermivelocity}) and (\ref{sommerfeld}) assuming the universal $v_{\rm F}$ inferred from photoemission measurements~\cite{horio2018,michon2019,hopping,drozdov2018}. Note that while the hole doping $p^\ast\approx$~0.24 at which the normal state pseudogap has been reported to close coincides with a Van Hove singularity, a finite interlayer hopping inferred from photoemission measurements has been argued to cause the calculated values of $\gamma$ at the Van Hove singularity to be truncated (blue dotted line)~\cite{horio2018}. Quasiparticle scattering effects have been argued to further erode the Van Hove singularity~\cite{michon2019,horio2018}. 

What we find to be particularly striking from the entropy data in Fig.~\ref{entropy} is that the measured values at room temperature are already between 20\% and 30\% of the fully saturated entropy of $R\ln4$  for a half-filled electronic band. As illustrated in Fig.~\ref{bandentropy}, an electronic entropy that rises more quickly with temperature than expected is a clear indication of a significant part of the spectral weight in the electronic density of states having been shifted to lower energies by correlation effects~\cite{georges1996}. 
One way these results can be understood is by considering a Gutzwiller-like renormalization of the quasiparticle dispersion analogous to that considered in $^3$He~\cite{vollhardt1984}. From the significant fraction of $R\ln4$ and the continued presence of the large slopes in $S$ versus $T$ at room temperature in Fig.~\ref{entropy}, we can infer that the strongly renormalized portion of the quasiparticle bands must extend over a range in energy at least of order $\sim$~30~meV on either side of the chemical potential.

We find that all of the thermodynamic data in Figs.~\ref{entropy}, \ref{orbitalvelocity} and \ref{lowtcalorimetry} can be understood by considering a significantly reduced hopping of $t\approx$~107~$\pm$~10~meV, corresponding to a reduced nodal Fermi velocity of $v_{\rm F}=$~(1.5~$\pm$~0.2)~$\times$~10$^5$~ms$^{-1}$. Here, $v_{\rm F}=$~1.5~$\times$~10$^5$~ms$^{-1}$ is the value we obtain by taking an average of the values of $v_{\rm F}$ that are able to produce entropy curves (red curves) that overlap with the experimental data (black curves) for each of the cuprates in Fig.~\ref{entropy}. 

In Fig.~\ref{orbitalvelocity}, we find that the reduced Fermi velocity of $v_{\rm F}=$~1.5~$\times$~10$^5$~ms$^{-1}$ (red horizontal line) is in quantitative agreement with $v^{\rm orb}_{\rm F}$ measured in the overdoped cuprate Tl2212~\cite{rourke2010}. It also provides an upper bound for $v^{\rm orb}_{\rm F}$ values measured in the underdoped regime ($p\lesssim$~0.2), where the Fermi surface is reconstructed.

In Fig.~\ref{lowtcalorimetry}a, we find that $t=$~107~meV is able to produce a peak in $\gamma$ versus $p$ (red curve) that is consistent with the experimentally measured values of $\gamma$ at low temperatures for $p\gtrsim$~0.24. For $p\lesssim$~0.24, the calculated values of $\gamma$ exceed those observed experimentally. However, $p\lesssim$~0.24 is also where the pseudogap opens, and this is also accompanied by long-range stripe ordering~\cite{ma2021,collignon2017}; we make no attempt to include these effects in Equation~(\ref{lowtcalorimetry}). 

Finally, in Fig.~\ref{lowtcalorimetry}b we find that insertion of $t=$~107~meV into Equation~(\ref{sommerfeld}) produces a logarithmically diverging $\gamma$ at $p=$~0.24 whose slope in $\ln T$ matches that measured in Eu-LSCO (and Nd-LSCO) over a broad span in temperature~\cite{michon2019}; this is a simple consequence of the Van Hove singularity. For the other compositions that were measured at $p<$~0.24~\cite{michon2019}, we find Equation~(\ref{sommerfeld}) to yield values of $\gamma$ that are saturated at near constant values  below 10~K, which is consistent with the rapid loss of $\ln T$ behavior once $p\neq$~0.24~\cite{weak}. 

The reduced $t$ implies an increased relevance of the Van Hove singularity in the phase diagram. While quantum criticality has been indicated as an important factor at $p^\ast\approx$~0.23 in Eu-LSCO and Nd-LSCO~\cite{legros2019}, we find that the majority of the increase in $\gamma$ is caused by the increased density of states of the Van Hove singularity (see Fig.~\ref{bandentropy}). Thus, there is little to no additional renormalization of $\gamma$ in low temperature calorimetry measurements (in Fig.~\ref{lowtcalorimetry}) beyond that simply accounted for by using the smaller values of $t\approx$~107~meV estimated from entropy measurements~\cite{loram2001,loram1994,loram1998}. 

We further note that whilst quantum criticality has been advocated as an explanation for the logarithmic divergence in Fig.~\ref{lowtcalorimetry}b, the agreement between Equation~(\ref{sommerfeld}) and the experimental slope in $\gamma$ versus $\ln{T}$ implies that interlayer tunneling (and scattering) cannot be suppressing the Van Hove singularity to the extent that has been suggested~\cite{tunnelingreason,hossain2010,horio2018,markiewicz1997}. A very recent study of LSCO  provides supporting arguments~\cite{zhong2022} for both the interlayer hopping and the effects of scattering having been overestimated in low temperature calorimetry measurements~\cite{michon2019}. The increased density of states of the Van Hove singularity caused by the smaller $t$ must therefore cause it to feature prominently in determinations of the quasiparticle scattering rate within the normal state at $p^\ast$~\cite{pattnaik1992,newns1994,legros2019}.

The reduced $t$ also implies that $m^\ast/m_{\rm b}=$~3.4~$\pm$~0.3. This renormalization lies between that  $\approx$~2.8 of ambient pressure $^3$He~\cite{vollhardt1984} and that $\approx$~4.0 of Sr$_2$RuO$_4$~\cite{bergemann2003} (taking an average over its three Fermi surface sheets), against which the cuprates are often compared~\cite{lee2006,bergemann2003}. A possible clue as to the origin of the difference between thermodynamic measurements and universal velocity inferred from photoemission measurements is provided by recent high resolution laser photoemission experiments, which find evidence for band flattening within $\omega\approx$~10~meV of the Fermi surface~\cite{vishik2010,plumb2010,wreedhar2020} (where $\omega=|\varepsilon-\mu|$). Many of the values of the nodal Fermi velocity obtained from laser photoemission ($\diamond$ symbols in Fig.~\ref{orbitalvelocity}) are found to be in excellent agreement with the value $v_{\rm F}=$~1.5~$\times$~10$^5$~ms$^{-1}$ we obtain from thermodynamic measurements. Laser photoemission and thermodynamic measurements are therefore in agreement on the existence of quasiparticles residing within $\sim$~10~meV of the Fermi surface. However, since a band flattening restricted to $\omega<$~10~meV cannot account for the excess entropy extending to high temperatures in Figs.~\ref{entropy} and~\ref{lowtcalorimetry}, it implies that a significant fraction of the excitations contributing to thermodynamic measurements at $\omega\gtrsim$~10~meV must be incoherent in photoemission measurements.  
One possibility is that coherent quasiparticles contributing to thermodynamic measurements constitute novel forms of composite or fractionalized fermion for which the corresponding features in photoemission measurements are weak or heavily broadened~\cite{jain2009,coleman2001,punk2015,moon2011,rice2012}. 

Finally, the reduced $t$ implies that $k_{\rm B}T_{\rm c}/t\approx$~0.1, which means that the pairing is sufficiently strong for $T_{\rm c}$ to be in a regime where phase fluctuations start to become an important consideration~\cite{hazra2019,shi2022,baskaran1987,emery1995}. Because the maximum theoretical ratio $k_{\rm B}T_{\rm c}/t\approx$~0.2~\cite{denteneer1993,chen1999,keller2001,toschi2005,paiva2010,pasrija2016,paiva2004} is based on the superfluid density of the large unreconstructed Fermi surface, a reduction in the superfluid density of the magnitude  reported by Uemura {\it et al.}~\cite{uemura1991,uemura1989} will be more than sufficient to give rise to a Brezinksii-Kosterlitz-Thouless transition or drive the superconductivity into the Bose pairing regime~\cite{shi2022,quasi2D}. Possible causes include the proximity to an insulating state~\cite{rullieralbenque2008} or the reconstruction of the Fermi surface into small pockets~\cite{ramshaw2015,tan2015,barisic2013,chan2016,kunisada2020,oliviero2022}.

\begin{acknowledgments}
The reduced Fermi velocity analysis was supported by the Department of Energy (DoE) BES project `Science of 100 tesla.' The entropy analysis was performed as part of a LDRD DR project at Los Alamos National Laboratory. The National High Magnetic Field Laboratory is funded by NSF Cooperative Agreements DMR-1157490 and 1164477, the State of Florida and DoE. NH thanks Arkady Shekhter for stimulating discussions. MKC acknowledges support from NSF IR/D program while serving at the National Science Foundation. Any opinion, findings, and conclusions or recommendations expressed in this material are those of the author(s) and do not necessarily reflect the views of the National Science Foundation.

\end{acknowledgments}

\bibliographystyle{naturemag}

\bibliography{basename of .bib file}

\begin{thebibliography}{99}

\bibitem{cao2018} Cao, Y., Fatemi, V., Fang, S., Watanabe, K., Taniguchi, T., Takashi, E. and Jarillo-Herrero, P., Unconventional superconductivity in magic-angle graphene superlattices. {\it Nature} {\bf 556}, 43-50 (2018).

\bibitem{park2021} Park, J.~M., Cao, Y., Watanabe, K., Taniguchi, T., Jarillo-Herrero, P., Tunable strongly coupled superconductivity in magic-angle twisted trilayer graphene. {\it Nature} {\bf 590}, 249-255 (2021).

%

\bibitem{lee2006} Lee, P. A., Nagaosa, N., Wen, X.-G. Doping a Mott insulator: Physics of high-temperature superconductivity. {\it Rev. Mod. Phys.} {\bf 78}, 17-85 (2006).

\bibitem{keimer2015} Keimer, B.,  Kivelson, S. A., Norman, M. R., Uchida, S. Zaanen, J. From quantum matter to high-temperature superconductivity in copper oxides. {\it Nature} {\bf 518}, 179-186 (2015).

%

\bibitem{timusk1999} Timusk, T., Statt, B. The pseudogap in high-temperature superconductors: an experimental survey. {\it Rep. Prog. Phys.} {\bf 62}, 61-122 (1999).

\bibitem{hufner2008} H\"{u}fner, S., Hossain, M.A., Damascelli, A., Sawatsky, G.A., Two gaps make a high-temperature superconductor? {\it Rep. Prog. Phys.} {\bf 71}, 062501 (2008).

\bibitem{harrison2022} Harrison, N., and Chan, M. K., Magic gap ratio for optimally robust fermionic condensation and Its implications for high-$T_{\rm c}$ Superconductivity. {\it Phys. Rev. Lett.} {\bf 129}, 017001 (2022). 

%

\bibitem{baskaran1987} Baskaran, G., Zou, Z. and Anderson, P. W., The resonating valence bond state and high-$T_{\rm c}$ superconductivity -- a mean field theory. {\it Solid State Commun.} {bf 63}, 973 (1987). 

\bibitem{uemura1989} Uemura, Y. J., Luke, G. M., Sternlieb, Brewer, J. H., Carolan, J. F., Hardy, W. N., Kadono, R., Kempton, J. R., Kiefl, R. F., Kreitzman, S. R., Mulhern, P., Riseman, T. M., Williams, D. L., Yang, B. X., Uchida, S., Takagi, H., Gopalakrishnan, J., Sleight, A. W., Subamanian, M. A., Chien, C. L., Cieolak, M. Z., Xiao, G., Lee, V. Y., Statt, N. W., Stronach, C. E., Kossler, W. J., Yu, X. H., Universal correlations between $T_{\rm c}$ and $n_{\rm s}/m^\ast$ (carrier density over effective mass) in high-$T_{\rm c}$ cuprate superconductors. {\it Phys. Rev. Lett.} {\bf 62}, 2317-2320 (1989).

\bibitem{uemura1991} Uemura, Y. J., Le, L. P., Luke, G. M., Sternlieb, B. J., Wu, W. D., Brewer, J. H., Riseman, T. M., Seaman, C. L., Maple, M. B., Ishikawa, M., Hinks, D. G., Jorgensen, J. D., Saito, G. and Yamochi, H. Basic similarities among cuprate, bismuthate, organic, Chevrel-phase, and heavy-fermion superconductors shown by penetration-depth measurements. {\it Phys. Rev. Lett.} {\bf 66}, 2665 (1991). 

\bibitem{emery1995} Emery, V.J., Kivelson, S.A., Importance of phase fluctuations in superconductors with small superfluid density. {\it Nature} {\bf 374}, 434-437 (1995).

\bibitem{hazra2019} Hazra, T., Verma, N. and Randeria, M., Bounds on the superconducting transition temperature: applications to twisted bilayer graphene and cold atoms. {\it Physical Review X} {\bf 9}, 031049 (2019).

\bibitem{shi2022} Shi, T. T., Zhang, W. and de Melo, C. A. R. S., Density-induced BCS-Bose evolution in gated two-dimensional superconductors: The role of the interaction range in the Berezinskii-Kosterlitz-Thouless transition. {\it EPL}, {\bf 139}, 36003 (2022).

\bibitem{quasi2D} The theories are mostly for a two-dimensional lattice for which a Brezinksii-Kosterlitz-Thouless (BKT) transition is expected. A BKT has been reported in the cuprates~\cite{li2007,guo2020}, while experiments on a quasi-two-dimensional system have reported the transition temperature to be only slightly higher than that of an ideal two-dimensional system~\cite{ries2015}.

\bibitem{li2007} Li, Q. Hucker, M., Gu, G. D., Tsvelik, A. M. and Tranquada, J. M., Two-dimensional superconducting fluctuations in stripe-ordered La$_{1.875}$Ba$_{0.125}$CuO$_4$. {\it Phys. Rev. Lett.} {\bf 99}, 067001 (2007). 

\bibitem{guo2020} Guo, J., Zhou, Y.-Z., Huang, C., Cai, S., Sheng, Y. T., Gu, G. D., Yang, C. L., Lin, G. C., Yang, K. 

\bibitem{ries2015} Ries, M. G., Wenz, A. N., Z\"{u}rn, G., Bayha, L., Boettcher, I., Kedar, D., Murthy, P. A., Neidig, M., Lompe, T. and Jochim, S. Observation of pair condensation in the quasi-2D BEC-BCS crossover. {\it Phys. Rev. Lett.} {\bf 114}, 230401 (2015). 

%


\bibitem{denteneer1993} Denteneer, P. J. H., An, G. and van Leeuwen, J. M. J., Helicity modulus in the two-dimensional Hubbard model. {\it Phys. Rev. B} {\bf 47}, 6256-7272 (1993). 

\bibitem{chen1999} Chen, Q., Kosztin, I., Jank\'{o}, B. and Levin, K. Superconducting transitions from the pseudogap state: $d$-wave symmetry, lattice, and low-dimensional effects. {\it Phys. Rev. B} {\bf 59}, 7083-7093 (1999).

\bibitem{keller2001} Keller, M., Metzner, W. and Schollw\"{o}ck, U., {\it Phys. Rev. Lett.} {\bf 86}, 4612-4615 (2001). 

\bibitem{paiva2004} Paiva, T., dos Santos, R. R., Scalettar, R. T. and Denteneer, P. J. H., Critical temperature for the two-dimensional attractive Hubbard model. {\it Phys. Rev. B} {\bf 69}, 184501 (2004). 

\bibitem{toschi2005} Toschi, A., Barone, P., Capone, M. and Castellani, C., Pairing and superconductivity from weak to strong coupling in the attractive Hubbard model. {\it New J. Phys.} {\bf 7}, 7 (2005).

\bibitem{paiva2010} Paiva, T., Scalettar, R., Randeria, M. and Trivedi, N., Fermions in 2D optical lattices: temperature and entropy scales for observing antiferromagnetism and superfluidity. {\it Phys. Rev. Lett.} {\bf 104}, 066406 (2010). 

\bibitem{pasrija2016}Pasrija, K., Chakraborty, P. B. and Kumar, S., Effective Hamiltonian based Monte Carlo for the BCS to BEC crossover in the attractive Hubbard model. {\it Phys. Rev. B} {\bf 94} 165150 (2016).

%

\bibitem{matteiss1987} Mattheiss, L. F. Electronic band properties and superconductivity in La$_{2-y}X_y$CuO$_4$. {\it Phys. Rev,. Lett.} {\bf 58}, 1028 (1987). 

\bibitem{massidda1988} Massidda S., Yu, J. and Freeman, A. J. Electronic structure and properties of Bi$_2$Sr$_2$CaCu$_2$O$_8$, the third high-$T_{\rm c}$ superconductor. {\it Physica C} {\bf 152}, 251 (1988). 

\bibitem{singh1992} Singh, D. J. and Pickett, W. E. Electronic characteristics of Tl$_2$Ba$_2$CuO$_6$. Fermi surface, positron wavefunction, electric field gradients and transport parameters. {\it Physica C} {\bf 203}, 193 (1992). 

\bibitem{andersen1995} Andersen, O. K., Liechtenstein, A. I., Jepsen, O. and Paulsen, E. LDA energy bands, low-energy Hamiltonians, $t^\prime$, $t^{\prime\prime}$, $t_\perp(k)$, and $J_\perp$. {\it J. Phys. Chem. Solids} {\bf 56}, 1573 (1995). 

%

\bibitem{lanzara2001}  Lanzara, A., Bogdanov, P. V., Zhou, X. J., Kellar, S. A., Feng, D. L.,  Lu, E. D., Yoshida, T., Eisaki, H., Fujimori, A., Kishio, K.,  Shimoyama, J.-I., Noda, T., Uchida, S., Hussain Z. and Shen Z.-X., Evidence for ubiquitous strong electron–phonon coupling in high-temperature superconductors. {\it Nature} {\bf 412}, 510 (2001). 

\bibitem{johnson2001} Johnson, P. D., Valla, T., Fedorov, A. V., Yusof, Z. , Wells, B. O., Li, Q., Moodenbaugh, A. R., Gu, G. D., Koshizuka, N., Kendziora, C., Jian, S. and Jinks, D. G.., Doping and temperature dependence of the mass enhancement observed in the Cuprate Bi$_2$Sr$_2$CaCu$_2$O$_{8-\delta}$. {\it Phys. Rev. Lett.} {\bf }, (2001).

\bibitem{byczuki2007} Byczuki, K., Kollar, M., Held, K., Yang, Y.-F., Nekrasov, I. A., Pruschke, TH. and Vollhardt, D. Kinks in the dispersion of strongly correlated electrons. {\it Nature Phys.} {\bf 3}, 168 (2007). 

%

\bibitem{padilla2005} Padilla, W. J.,  Lee, Y. S., Dumm, M., Blumberg, G., Ono, S., Segawa, K.,  Komiya, S., Ando Y. and  Basov, D. N., Constant effective mass across the phase diagram of high-$T_{\rm c}$ cuprates. {\it Phys. Rev. B} {\bf 72}, 060511(R) (2005). 

\bibitem{doiron2007} Doiron-Leyraud, N., Proust, C. LeBoeuf, D., Levallois, J., Bonnemaison, B., Liang, R.-X. Hardy, W. N., Bonn, D. A., Taillefer, L. Quantum oscillations and the Fermi surface in an underdoped high-$T_{\rm c}$ superconductor. {\it Nature} {\bf 447}, 565-568 (2007).

\bibitem{storey2008} Storey, J. G., Tallon, J. L., and Williams, G. V. M., Thermodynamic properties of Bi$_2$Sr$_2$CaCu$_2$O$_8$ calculated from the electronic dispersion. {\it Phys. Rev. B} {\bf 77}, 052504 (2008).

\bibitem{sebastian2012} Sebastian, S.E., Harrison, N., Lonzarich, G.G. Towards resolution of the Fermi surface in underdoped high-$T_{\rm c}$ superconductors. {\it Rep. Prog. Phys.} {\bf 75}, 102501 (2012).

\bibitem{ramshaw2015} Ramshaw, B. J., Sebastian, S. E., McDonald, R. D.,  Day, J., Tan, B. S., Zhu, Z., Betts, J.B., Liang, R.-X., Bonn, D.A., Hardy, W.N., Harrison, N., Quasiparticle mass enhancement approaching optimal doping in a high-$T_{\rm c}$ superconductor. {\it Science} {\bf 348}, 317-320 (2015).

\bibitem{michon2019} Michon, B., Girod, C., Badoux, S., Ka\u{c}mar\u{c}\'{i}k, J., Ma, Q., Dragomir, M., Dabkowska, H. A., Gaulin, B.D., Zhou, J.-S., Pyon, S., Takayama, T., Takagi, H., Verret, S., Doiron-Leyraud, N., Marcenat, C., Taillefer, L., Klein, T., Thermodynamic signatures of quantum criticality in cuprate superconductors. {\it Nature} {\bf 567}, 218-222 (2019). 

\bibitem{legros2022} Legros, A., Post, K. W., Chauhan, P., Rickel, D. G., He, X., Xu, X., Shi, X., B\u{o}zovi\'{c}, I., Crooker, S. A. and Armitage, N. P., Evolution of the cyclotron mass with doping in La$_{2-x}$Sr$_x$CuO$_4$. {\it Phys. Rev. B} {\bf 106}, 195110 (2022).

\bibitem{zhong2022} Zhong, Y., Chen, Z., Chen, S.-D., Xu, K.-J., Hashimoto, N., He, Y., Mo, S.-K. and Shen, Z.-X., Differentiated roles of Lifshitz transition on thermodynamics and superconductivity in La$_{2-x}$Sr$_x$CuO$_4$. {\it Proc. Natl. Acad. USA} {\bf 119}, e2204630119 (2022). 

\bibitem{sous2022} Sous, J., He, Y. and Kivelson, S. A., Absence of a BCS-BEC crossover in the cuprate superconductors. arXiv:cond-mat/0610442 (2022).

%

\bibitem{yoshida2003} Zhou, X. J., Yoshida, T., Lanzara, A., Bogdanov, P. V., Kellar, S. A., Shen, K. M., Yang, W. L., Ronning, F., Sasagawa, T., Kakeshita, T., Noda, T., Eisaki, H., Uchida, S., Lin, C. T., Zhou, F., Xiong, J. W., Ti, W. X., Zhao, Z. X., Fujimori, A., Hussain, Z. and Shen, Z.-X., Universal nodal Fermi velocity. {\it Nature} {\bf 423}, 398 (2003). 

\bibitem{damascelli2003} Damascelli, A., Hussain, Z., Shen, Z.-X. Angle-resolved photoemission studies of the cuprate superconductors. {\it Rev. Mod. Phys.} {\bf 75}, 473-541 (2003).


%
%


%

\bibitem{horio2018} Horio, M., Hauser, K., Sassa, Y., Mingazheva, Z., Sutter, D., Kramer, K., Cook, A., Nocerino, E.,
Forslund, O. K., Tjernberg, O., Kobayashi, M., Chikina, A., Schr\"{o}ter, N. B. M., Krieger, J. A., 
Schmitt, T., Strocov, V. N., Pyon, S., Takayama, T., Takagi, H., Lipscombe, O. J., Hayden, S. M., Ishikado, M., Eisaki, H., Neupert, T.,  M\o{a}nsson, M., Matt, C. E. and Chang, J. Three-dimensional Fermi surface of overdoped La-based cuprates. {\it Phys. Rev. Lett.} {\bf 121}, 077004 (2018). 

\bibitem{hopping} For tight-binding electronic band, the nodal Fermi velocity at ${\bf k}=(\pi/2a,\pi/2a)$ is given by $v_{\rm F}=2\sqrt{2}at/\hbar$, where $a$ is the in-plane lattice parameter, for which we obtain $t\approx$~164~meV. However, because the Fermi surface does not coincide with ${\bf k}=(\pi/2a,\pi/2a)$~\cite{drozdov2018}, the hopping required to produce the experimental value is 190~meV~\cite{horio2018,michon2019}. 

\bibitem{drozdov2018} Drozdov, I . K., Pletikosi\'{c}, I., Kim, C.-K., Fujita, K., Gu, G. D., Davis, J. C. S., Johnson, P. D., Bo\u{z}ovi\'{c}, I. and Valla, T., Phase diagram of Bi$_2$Sr$_2$CaCu$_2$O$_{8+\delta}$ revisited. {\it Nature Commun.} {\bf 9}, 5210 (2018). 

%

\bibitem{mathur1998} Mathur, N. D., Grosche, F. M., Julian, S. R., Walker,, I. R., Freye, D. M., Haselwimmer, R. K. W. and Lonzarich, G. G., Magnetically mediated superconductivity in heavy fermion compounds. {\it Nature} {\bf 394}, 39-43 (1998). 

\bibitem{laughlin2001} Laughlin, R. B., Lonzarich, G. G., Monthoux, P. and Pines, D. The quantum criticality conundrum. {\it Adv. Phys.} {\bf 50}, 361-365 (2001)

\bibitem{sachdev2010} Sacdev, S. Where is the quantum critical point int he cuprate superconductors? {\it Phys. Status Solidi} {\bf B247}, 537-543 (2010). 

\bibitem{shibauchi2013} Shibauchi, T., Carrington, A. and Matsuda, Y., A quantum critical point lying beneath the superconducting dome in iron pnictides. {\it  Annual Review of Condensed Matter Physics} {\bf 5}, 113-135 (2013). 

\bibitem{coleman2005} Coleman, P. and Schofield, A. J., {\it Nature} {\bf 433}, 226-229 (2005). 

%

\bibitem{loram1994} Loram, J. W., Mirza, K. A., Cooper, J. R. Liang, W. Y. and Wade, J. M., Electronic specific heat of YBa$_2$Cu$_3$0$_{6+x}$ from 1.8 to 300 K. {\it J. Superconductivity} {\bf 7}, 243-249 (1994).

\bibitem{loram1998} Loram, J. W., Mirza, K. A., Cooper, J. R., Tallon, J. L. Specific heat evidence of the normal state pseudogap. {\it J. Phys. Chem. Solids} {\bf 59}, 2091-2094 (1998).

\bibitem{loram2001} Loram, J. W., Luo, J., Cooper, J. R., Liang, W. Y., Tallon, J. L., Evidence on the pseudogap and condensate from the electronic specific heat, {\it J. Physics and Physical Chemistry of Solids} {\bf 62}, 59-64 (2001).

%

\bibitem{rourke2010} Rourke, P. M. C., Bangura, A. F., Benseman, T. M., Matusiak, M., Cooper, J. R., Carrington, A., Hussey, N. E., A detailed de Haas–van Alphen effect study of the overdoped cuprate Tl$_2$Ba$_2$CuO$_{6+\delta}$, {\it New J. Phys.} {\bf 12}, 105009 (2010).

\bibitem{barisic2013} Bari\u{s}i\'{c}, N., Badoux, S., Chan, M. K., Dorow, C., Tabis, W., Vignolle, B., Guichan, Y., Beard, J., Zhao, X., Proust, C. Universal quantum oscillations in the underdoped cuprate superconductors, {\it Nature Phys.} {\bf 9}, 761-764 (2013).

\bibitem{tan2015} Tan, B. S. , Harrison, N., Zhu, Z., Balakirev, F., Ramshaw, B. J., Srivastava, A., Sabok-Sayr, S. A., Dabrowski, B., Lonzarich, G. G. and Sebastian, S. E. Fragile charge order in the nonsuperconducting ground state of the underdoped high-temperature superconductors. {\it Proc. Natl. Acad. USA} {\bf 112} 9568 (2015). 

\bibitem{chan2016} Chan, M. K., Harrison, N., McDonald, R. D., Ramshaw, B. J., Modic, K. A., Bari\u{s}i\'{c}, N. and Greven, M., Single reconstructed Fermi surface pocket in an underdoped single-layer cuprate superconductor. {\it Nature Commun.} {\bf 7}, 12244 (2016). 

\bibitem{kunisada2020} Kunisada, S., Isono, S., Kohama, Y., Sakai, S., Bareille, C., Sakuragi, S., Noguchi, R., Kurokawa, K., Kuroda, K., Ishida, Y., Adachi, S., Sekine, R., Kim, T. K., Cacho, C., Shin, S., Tohyama, T., Tokiwa, K., Kondo, Y., Observation of small Fermi pockets protected by clean CuO$_2$ sheets of a high-$T_{\rm c}$ superconductor. {\it Science} {\bf 369}, 833-838 (2020).

\bibitem{oliviero2022} Oliviero, V., Benhabib, S., Gilmutdinov, I., Vignolle, B., Drigo, L., Massoudzadegan, M.,  Leroux, M., Rikken, G. L. J. A., Forget, A., Colson, D., Vignolles, D. and Proust, C. Magnetotransport signatures of antiferromagnetism coexisting with charge order in the trilayer cuprate HgBa$_2$Ca$_2$Cu$_3$O$_{8+\delta}$. {\it Nature Commun.} {\bf 13}, 1568 (2022).

%

\bibitem{tightbindingparameters} We have used $\epsilon_{\bf k}=-2t(\cos k_x+\cos k_y)+4t^\prime\cos k_x\cos k_y-2t^{\prime\prime}(\cos2k_x+\cos2k_y)-\mu$ for the tight-binding dispersion where we have used $t^\prime/t=$~$-$~0.32, $-$~0.12, $-$~0.30, $-$~0.32 and $-$~0.14, and $t^{\prime\prime}/t^\prime=$~$-$~0.50, $-$~0.50, $-$~0.33, $-$~0.5, and $-$~0.5, for YBCO, LSCO Bi2210, Ca-YBCO, and Nd- and Eu-LSCO, respectively~\cite{matteiss1987,andersen1995,drozdov2018,horio2018}. The electronic density of states is obtained using $D(\varepsilon_{\bf k})=\frac{n}{\pi^2}\frac{\partial}{\partial\varepsilon_{\bf k}}\int^\pi_0k_y(\varepsilon_{\bf k}){\rm d}k_x$.

%

\bibitem{ashcroft1976} Ashcroft, N. W. and Mermin, N. D. {\it Solid state physics} (Saunders College Publishing, Orlando 1976).

%

\bibitem{harrison2022i} Wartenbe, M., Tobash, P. H., Singleton, J., Winter, L. E., Richmond, S. and Harrison, N., Pseudogap in elemental plutonium. {\it Phys. Rev. B} {\bf 105}, L041107 (2022).

%

\bibitem{shoenberg1984} Shoenberg, D. {\it Magnetic Oscillations in Metals} (Cambridge Univ. Press, 1984).

%

\bibitem{otherfactors} Other factors could include a Lifshitz transition, a Van Hove singularity occurring at a different doping in other cuprates, or quantum criticality associated with the phase that reconstructs the Fermi surface. 

%

\bibitem{georges1996} Georges, A., Kotliar, G., Krauth, W. and Rozenberg, M. J., Dynamical mean-field theory of strongly correlated fermion systems and the limit of infinite dimensions. {\it Rev. Mod. Phys.} {\bf 68}, 13-125 (1996).

%

\bibitem{vollhardt1984} Vollhardt, D., Normal $^3$He: an almost localized Fermi liquid. {\it Rev. Mod. Phys.} {\bf 56}, 99 (1984).

%

\bibitem{ma2021} Ma, Q., Rule, K. C., Cronkwright, Z. W., Dragomir, M., Mitchell, G., Smith, E. M., Chi, S., Kolesnikov, A. I., Stone, M. B., Gaulin, B. D., Parallel spin stripes and their coexistence with superconducting ground states at optimal and high doping in La$_{1.6-x}$Nd$_{0.4}$Sr$_x$CuO$_4$. {\it Phys. Rev. Research} {\bf 3}, 023151 (2021).

\bibitem{collignon2017} Collignon, C., Badoux, S., Afshar, S. A. A., Michon, B.,  Lalibert\'{e}, F., Cyr-Choini\`{e}re, O., Zhou, J.-S., Licciardello, S., Wiedmann, S., Doiron-Leyraud, N. and Taillefer, L. Fermi-surface transformation across the pseudogap critical point of the cuprate superconductor La$_{1.6-x}$Nd$_{0.4}$Sr$_x$CuO$_4$. {\it Phys. Rev. B} {\bf 95}, 224517 (2017).

%

\bibitem{weak} We speculate that residual weaker slopes in $\ln T$ at $p=$~0.21 and $p=$~0.22 could be the consequence of slight sample inhomogeneity. In our calculations, we find that for $p=$~0.21 and $p=$~0.22, $\gamma$ versus $T$ exhibits a kink at $T>$~10~K, above which the slope in $\ln T$ is similar to that at $p=$~0.24. However, there is no experimental data in this region. Finally, $p<$~0.24 is where the pseudogap is argued to open, whose effect in reducing $\gamma$ we make no attempt to model.

%

\bibitem{tunnelingreason} This could be a consequence of the assumed form of the interlayer dispersion being poorly representative of that in experiments~\cite{horio2018}, the presence of higher harmonics in the hopping, or a suppression of the interlayer hopping caused by electronic correlations or quantum criticality~\cite{hossain2010}. Other photoemission measurements, for example, have indicated that the Van~Hove singularity is more extended in momentum-space~\cite{markiewicz1997}.

\bibitem{hossain2010} Fournier, D., Levy, G., Pennec, Y., McChesney, J. L., Bostwick, A., Rotenberg, E., Liang, R., Hardy, W. N., Bonn, D. A., Elfimov, I. S. and Damascelli, A., Loss of nodal quasiparticle integrity in underdoped YBa$_2$Cu$_3$O$_{6+x}$. {\it Nature Phys.} {\bf 6}, 905-911 (2010).

\bibitem{markiewicz1997} Markiewicz, R. S., A Survey of the Van Hove scenario for high-$T_{\rm c}$ superconductivity with special emphasis on pseudogaps and striped phases. {\it J. Phys. Cha,. Solids} {\bf 58}, 1179-1310 (1997).

%

\bibitem{legros2019} Legros, A., Benhabib, S., Tabis, W., Lalibert\'{e}, F., Dion, M., Lizaire, M., Vignolle, B., Vignolles, D., Raffy, H., Li, Z. Z., Auban-Senzier, P., Doiron-Leyraud, N., Fournier, P., Colson, D., Taillefer, L. and Proust, C., Universal $T$-linear resistivity and Planckian dissipation in overdoped cuprates. {\it Nature Phys.} {\bf 15}, 142-147 (2019). 

%

\bibitem{pattnaik1992} Pattnaik, P. C., Kane, C. L., Newns, D. M. and Tsuei, C. C., Evidence for the van Hove scenario in high-temperature superconductivity from quasiparticle-lifetime broadening. {\it Phys. Rev. B} {\bf 45}, 5714-5717 (1992). 

\bibitem{newns1994} Newns, D. M., Tsuei, C. C., Huebener, R. P., van Bentum, P. J. M., Pattnaik, P. C. and Chi, C. C. Quasiclassical Transport at a van Hove Singularity in Cuprate Superconductors. {\it Phys. Rev. Lett.} {\bf 73}, 1695-1698 (1994). 

%

\bibitem{bergemann2003} Bergemann, C., Mackenzie, A. P., Julian, S. R., Forsythe, D. and Ohmichi, E., Quasi-two-dimensional Fermi liquid properties of the unconventional superconductor Sr$_2$RuO$_4$. {\it Adv. Phys.} {\bf 52}, 639 (2003).

%

\bibitem{vishik2010} Vishik, I. M., Lee, W. S., Schmitt, F., Moritz, B., Sasagawa, T., Uchida, S., Fujita, K., Ishida, S., Zhang, C., Devereaux, T. P. and Shen, Z. X., Doping-dependent nodal Fermi velocity of the high-yemperature superconductor Bi$_2$Sr$_2$CaCu$_2$O$_{8+\delta}$ revealed using high-resolution angle-resolved
photoemission spectroscopy. {\it Phys. Rev. Lett.} {\bf 104}, 207002 (2010). 

\bibitem{plumb2010} Plumb, N. C., Reber, T. J., Koralek, J. D., Sun, Z., Douglas, J. F., Aiura, Y., Oka, K., Eisaki, H. and Dessau, D. S., Low-energy ($<$10 meV) feature in the nodal electron self-energy and strong temperature
dependence of the Fermi velocity in Bi$_2$Sr$_2$CaCu$_2$O$_{8+\delta}$. {\it Phys. Rev. Lett.} {\bf 105}, 046402 (2010). 

\bibitem{wreedhar2020} Sreedhar, S. A., Rossi, A., Nayak, J., Anderson, Z. W., Tang, Y., Gregory, B., Hashimoto, M., Lu, D.-H., Rotenberg, E., Birgeneau, R. J., Greven, M., Yi, M. and Vishik, I. M., Three interaction energy scales in the single-layer high-$T_{\rm c}$ cuprate HgBa$_2$CuO$_{4+\delta}$. {\it Phys. Rev. B} {\bf 102}, 205109 (2020).

%

\bibitem{jain2009} Jain, J. K., and Anderson, P. W. Beyond the Fermi liquid paradigm: hidden Fermi liquids. {\it Proc. Natl. Acad. Sci. USA} {\bf 106}, 9131-9134 (2009).
%
\bibitem{coleman2001} Coleman, P., P\'{e}pin, C., Si, Q., Ramazashvili, R. How do Fermi liquids get heavy and die? {\it J. Phys.: Condens. Matter} {\bf 13}, R723 (2001).
%
\bibitem{punk2015} Punk, M., Allais, A. and Sachdev, S., Quantum dimer model for the pseudogap metal. {\it Proc. Natl. Acad. USA} {\bf 112}, 9552-9557 (2015).
%
%
\bibitem{moon2011} Moon, E. G., and Sachdev, S., Underdoped cuprates as fractionalized Fermi liquids: Transition to superconductivity. {\it Phys. Rev. B} {\bf 83}, 224508 (2011).
%
\bibitem{rice2012} Rice, T.M., Yang, K.-Y., Zhang, F.C., A phenomenological theory of the anomalous pseudogap phase in underdoped cuprates. {\it Rep. Prog. Phys.} {\bf 75}, 016502 (2012).



%
%
%
%
%

%

%
%

%


%


%
%
%
%
%
%

%


%


%



%


%

\bibitem{rullieralbenque2008} Rullier-Albenque, F., Alloul, H., Balakirev, F., Proust, C., Disorder, metal-insulator crossover and phase diagram in high-$T_{\rm c}$ cuprates. {\it EPL} {\bf 81}, 37008 (2008).





%


%






%





%
%
%
%
%
%
%
%
%
%
%
%
%
%
%
%
%
%
%
%
%
%
%
%
%
%
%
%
%
%
%
%
%
%
%
%
%
%
%
%
%
%
%
%
%
%
%
%
%
%
%
%
%
%
%
%
%
%
%
%
%
%
%
%
%
%
%
%
%
%
%
%
%
%
%
%
%
%
%
%
%
%
%
%
%
%
%
%
%
%
%
%
%
%
%
%
%
%
%
%
%
%
%



%
%
%
%
%
%
%
%
%
%
%
%
%
%
%
%
%
%
%
%
%
%
%
%
%
%
%

%
%
%
%



%
%
%
%
%
%
%
%
%
%
%
%
%
%
%
%
%
%
%

%
%


%



%


%


%



%



%
%







%
%
%
%
%
%
%
%
%
%




%




%


%










%
%
%
%
%
%
%
%
%
%
%
%
%
%
%
%
%
%
%
%
%
%
%
%
%
%
%
%
%
%
%
%
%
%
%
%
%
%
%
%
%
%
%
%
%
%
%
%
%
%
%
%
%
%
%
%
%
%
%
%
%
%
%
%
%
%
%
%
%
%
%
%
%
%
%
%
%
%
%
%
%
%
%
%
%
%
%
%
%
%
%
%
%
%
%
%
%
%
%
%
%
%
%
%
%
%
%
%
%
%
%
%
%
%
%
%
%
%
%
%
%
%
%
%
%
%
%
%
%
%
%
%
%
%
%
%
%
%
%
%
%
%
%
%
%
%
%
%
%
%
%
%
%
%
%
%
%
%
%
%
%
%
%
%
%
%
%
%
%
%
%
%
%
%
%
%
%
%
%
%
%
%
%
%
%
%
%
%
%
%
%
%
%
%
%
%
%
%
%
%
%
%
%
%
%
%
%
%
%
%
%
%
%
%
%
%
%
%
%
%
%
%
%
%
%
%
%
%
%
%
%
%
%
%
%
%
%
%
%
%
%
%
%
%
%
%
%
%
%
%
%
%
%
%
%
%
%
%
%
%
%
%
%
%
%
%
%
%
%
%
%
%
%
%
%
%
%
%
%
%
%
%
%
%
%
%
%
%
%
%
%
%
%
%
%
%
%
%
%
%
%
%
%
%
%
%
%
%
%
%
%
%
%
%
%
%
%
%
%
%
%
%


%



\end{thebibliography}

\end{document}